\documentclass[twocolumn,superscriptaddress,letter]{revtex4}%
\usepackage{amssymb}
\usepackage{color}
\usepackage{graphicx}
\usepackage{dcolumn}
\usepackage{bm}
\usepackage[header,title,page,titletoc]{appendix}
\usepackage{amsmath}
\usepackage{subfigure}
\usepackage{amsfonts}%
\providecommand{\U}[1]{\protect \rule{.1in}{.1in}}

\begin{document}
\title{Verifying Non-Abelian Statistics by Numerical Braiding Majorana Fermions}
\author{Qiu-Bo Cheng}
\thanks{These authors contributed equally to this paper.}
\affiliation{Department of Physics, Beijing Normal University, Beijing, 100875, P. R. China}
\author{Jing He}
\thanks{These authors contributed equally to this paper.}
\affiliation{Department of Physics, Hebei Normal University, Hebei, 050024, P. R. China}
\author{Su-Peng Kou}
\thanks{Corresponding author}
\email{spkou@bnu.edu.cn}
\affiliation{Department of Physics, Beijing Normal University, Beijing, 100875, P. R. China}

\begin{abstract}
Recently, Majorana fermions (MFs) have attracted intensive attention because
of their possible non-Abelian statistics. This paper points out an approach to
verify the non-Abelian statistics of MFs in topological superconductors. We
introduce a single particle representation of braiding operators that obey
anti-commutating relation of Bogolubov-de Gennes (BdG) states. From the
relationship between the braiding operator of MFs and that of BdG states, we
verify non-Abelian statistics of MFs in 1D and 2D topological SCs.

\end{abstract}
\maketitle

Majorana fermions (MFs) $\gamma_{i}$ are their own antiparticle and constitute
`half' of ordinary fermions\cite{Majorana,Wilczek,Martin} that satisfies
$\gamma_{i}=\gamma_{i}^{\dag}$ and $\{ \gamma_{i},\gamma_{j}\}=2\delta_{ij}$.
It is still unclear whether MFs exist in nature as elementary building blocks,
but in condensed matter systems, MFs may appear as Majorana bound states
(MBSs)\cite{Kitaev01,Fu08,Lutchyn10, Oreg10, Sau10,Alicea10,Potter10,
Alicea11,Halperin12,Stanescu13,Mourik12,Das12,Deng12,Rokhinson12,Churchill13}.
Recently, due to their potential applications in topological quantum
computation (TQC)\cite{ki2,Nayak08,free,sar1,ge}, the search for exotic states
supporting MFs has attracted increasing interest in condensed matter physics.
Possible example of such quantum exotic states is the $p_{x}+ip_{y}$
topological superconductor (SC). The quantized vortex ($\pi$-flux) in two
dimensional (2D) $p_{x}+ip_{y}$ topological SCs trap
MBSs\cite{Read00,Ivanov01,Nayak08,Alicea12}. Another creative proposal is the
interface of $s$-wave SCs and topological insulators due to the proximity
effect\cite{Fu08}, in which a quantized vortex may also trap MBSs. In
addition, a new type of MBSs proposed by Kitaev occurs in a one-dimensional
(1D) electronic nanowire proximity-coupled to a bulk
superconductor\cite{Kitaev01}. For this case, two unpaired Majorana zero modes
appear at the two ends of the nanowire. In Ref.\cite{kou}, MBSs are also found
to be trapped by line defects of p-wave superconductors on a honeycomb lattice.

In Ref.\cite{Read00}, it was pointed out that the MFs trapped by vortices in
$p_{x}+ip_{y}$-wave topological SC obey non-Abelian statistics\cite{moo,wen}.
After exchanging two MFs in the 2D topological SC, the braiding operation can
be represented by $\gamma_{1}\rightarrow \gamma_{2},$ $\gamma_{2}%
\rightarrow-\gamma_{1}$\cite{Ivanov01}. For the MFs at the ends of line-defect
in SCs, the braiding operations were illustrated along T-junction
pathes\cite{Alicea11}. Based on the arguments in Ref.\cite{Alicea11}, the
line-defect-induced MFs were found to obey non-Abelian statistics. As a
result, the MFs in different models (both vortex-induced MFs and the
line-defect-induced MFs) were believed to obey the same type of non-Abelian
statistics. In this paper, we will introduce a numerical approach to verify
the statistics of MFs. The results will be helpful to learn the properties of
topological SCs.

In certain topological SCs, MFs with zero energy (the Majorana zero modes) may
emerge around topological defects (for example, the quantized vortex or the 1D
nano-wire). In general, people can obtain the function $\left \vert
\gamma \right \rangle $ of an \emph{emergent }Majorana zero mode by solving the
Bogolubov-de Gennes (BdG) equations numerically. To describe the Majorana zero
mode, a real fermion field called Majorana fermion $\gamma=\int d{r}[u_{0}%
\psi^{\ast}+v_{0}\psi]$ ($\gamma^{\dagger}=\gamma$) is introduced. To describe
the subspace of the system with two degenerate BdG states (Majorana zero
modes) $\Phi_{1}(x),$ $\Phi_{2}(x)$ (that correspond to two MFs $(\gamma_{1},$
$\gamma_{2})$), we introduce the Fermion-parity operator $\mathrm{\hat{P}%
}=(-i\gamma_{1}\gamma_{2})$. Since $\mathrm{\hat{P}}^{2}=1$, $\mathrm{\hat{P}%
}$ has two eigenvalues $\pm1,$ called even and odd Fermion-parities,
respectively. In a 2D gapped SC, the quantum eigen-states including Majorana
modes must have a determinant parity. Then, we label a pair of MFs
$(\gamma_{1},$ $\gamma_{2})$ by a complex fermion as $\gamma_{1}=c+c^{\dag},$
$\gamma_{2}=-i(c-c^{\dag})$ where $c^{\dagger}/c$ is the creation/annihilation
operator of spinless fermion. With the help of $c$ and $c^{\dag}$, two
Majoarna modes $(\gamma_{1},$ $\gamma_{2})$ form the physical Fermion-parity
qubit: $\left \vert 0\right \rangle $ is a many-body quantum state with even
Fermion-parity and $c^{\dag}\left \vert 0\right \rangle =\left \vert
1\right \rangle $ is a many-body quantum state with odd Fermion-parity.
Generally, there exists the coupling between two MFs and the effective
Hamiltonian is given by
\begin{equation}
i\mathcal{J}\gamma_{1}\gamma_{2}=2\mathcal{J}\left(  c^{\dag}c-\frac{1}%
{2}\right)
\end{equation}
where $\mathcal{J}$ is the coupling constant.

To distinguish the statistics for the MFs, we firstly show their braiding
operations in many-body representation. In quantum mechanics, there are two
possibilities of the \emph{many-body} wave-function to change by a $\pm$ sign
under a single particle-interchange (the so-called braiding operation in 2D
systems), corresponding to the cases of bosons and fermions, respectively.
Then after braiding two identical particles, the many-body wave-function
$\left \vert \psi_{\mathrm{initial}}\right \rangle =a^{\dagger}(\mathbf{r_{1}%
})a^{\dagger}(\mathbf{r_{2}})\left \vert 0\right \rangle $ changes by a phase to
be $\left \vert \psi_{\mathrm{initial}}\right \rangle \rightarrow \left \vert
\psi_{\mathrm{final}}\right \rangle =e^{i\theta}a^{\dagger}(\mathbf{r_{2}%
})a^{\dagger}(\mathbf{r_{1}})\left \vert 0\right \rangle .$ $\theta=0,\pi$
correspond to bosons and fermions, respectively.

However, because the emergent MFs are non-local and there is no generation
operator for MFs, people cannot do the braiding operations by local operators.
Instead, people do the braiding operation on MFs by adiabatically deforming
the system, for example, changing the length of a 1D nano-wire or moving the
vortices. That means \emph{the braiding operations for MFs are really
adiabatic evolutions} of the system. According to the results in
Ref.\cite{Ivanov01}, if we exchange two MFs with non-local $\pi$ phase
strings, the resulting braiding operation is given by
\begin{equation}
\gamma_{1}\rightarrow \gamma_{2},\gamma_{2}\rightarrow-\gamma_{1}\label{b-n}%
\end{equation}
that can be described by an "unitary" transformation $U_{ij}=e^{\frac{\pi}%
{4}\gamma_{i}\gamma_{j}}$\cite{Ivanov01}. For three MFs $\gamma_{i}$,
$\gamma_{j}$, $\gamma_{k}$, due to $\left[  U_{ij},\text{ }U_{jk}\right]
\neq0$ and $\{U_{ij},$ $U_{jk}\} \neq0,$ an adiabatic braiding operation
obviously shows a non-Abelian character of the MFs. On the basis of many-body
Majorana states $(%
\begin{array}
[c]{c}%
\left \vert 1\right \rangle \\
\left \vert 0\right \rangle
\end{array}
),$ the resulting braiding operator is $e^{-i\pi \sigma^{z}/4}$. Based on
braiding operations on Ising anyons that obey non-Abelian statistics, people
can do the Hadamard gate, the phase gate, and the $\mathrm{CNOT}$ gate
topologically except for the $\pi/8$ gate\cite{nay,free1,ge}.

Because all energy levels in the topological superconductors can be obtained
by numerical method accurately, we can also explore non-Abelian statistics of
MFs from BdG states in single particle representation. In particular, during
braiding operations, the adiabatic evolutions of the degenerate energy levels
(BdG states) with a pair of MFs $\Phi_{1}(x),$ $\Phi_{2}(x)$ lead to a
nontrivial change, $\left \vert \psi_{\mathrm{initial}}\right \rangle =(%
\begin{array}
[c]{c}%
\left \vert \gamma_{1}\right \rangle \\
\left \vert \gamma_{2}\right \rangle
\end{array}
)\rightarrow \left \vert \psi_{\mathrm{final}}\right \rangle =\mathcal{R}(%
\begin{array}
[c]{c}%
\left \vert \gamma_{1}\right \rangle \\
\left \vert \gamma_{2}\right \rangle
\end{array}
)=(%
\begin{array}
[c]{c}%
\left \vert \gamma_{1}^{\prime}\right \rangle \\
\left \vert \gamma_{2}^{\prime}\right \rangle
\end{array}
)$. Owning to the conservation rule of Fermion-parities, the mixing of
diagonalized BdG states $\left \vert +\right \rangle =(\left \vert \gamma
_{1}\right \rangle +i\left \vert \gamma_{2}\right \rangle )/\sqrt{2}$ and
$\left \vert -\right \rangle =(\left \vert \gamma_{1}\right \rangle -i\left \vert
\gamma_{2}\right \rangle )/\sqrt{2}$ is forbidden. Then due to the
particle-hole (PH) symmetry, the final BdG functions of the Majorana modes to
which the system returns after the braiding is identical to the initial one,
up to a phase,
\begin{equation}
\left \vert \psi_{\mathrm{initial}}\right \rangle =(%
\begin{array}
[c]{c}%
\left \vert +\right \rangle \\
\left \vert -\right \rangle
\end{array}
)\rightarrow \left \vert \psi_{\mathrm{final}}\right \rangle =(%
\begin{array}
[c]{c}%
e^{i(\varphi_{\mathrm{Dynamical}}+\varphi_{\mathrm{Berry}})}\left \vert
+\right \rangle \\
e^{-i(\varphi_{\mathrm{Dynamical}}+\varphi_{\mathrm{Berry}})}\left \vert
-\right \rangle
\end{array}
)
\end{equation}
where the Barry phase $\varphi_{\mathrm{Berry}}$ does not depend on how long
the process takes and the dynamical phase $\varphi_{\mathrm{Dynamical}}%
=\int_{0}^{\Delta \mathrm{t}}J(\mathrm{t})d\mathrm{t}$ depends on the energy of
the BdG states $J(\mathrm{t})$ and the length of time for the process
$\Delta \mathrm{t}$.

Let us derive their braiding operators for BdG states in single particle
representation. When we consider the initial BdG states $\left \vert
\psi_{\mathrm{i}}\right \rangle $ to be $(%
\begin{array}
[c]{c}%
\left \vert \gamma_{1}\right \rangle \\
\left \vert \gamma_{2}\right \rangle
\end{array}
)$, after the braiding operation, $2\longleftrightarrow1$, the final BdG
states $\left \vert \psi_{\mathrm{f}}\right \rangle $ turn into $(%
\begin{array}
[c]{c}%
e^{i\phi_{1}}\left \vert \gamma_{2}\right \rangle \\
e^{i\phi_{2}}\left \vert \gamma_{1}\right \rangle
\end{array}
),$ where $e^{i\phi_{1}}$ and $e^{i\phi_{2}}$ are phase factors to be solved.
On the other hand, when we consider the initial BdG states $\left \vert
\psi_{\mathrm{i}}\right \rangle $ to be $(%
\begin{array}
[c]{c}%
\left \vert +\right \rangle \\
\left \vert -\right \rangle
\end{array}
)$, after the braiding operation, $+\longleftrightarrow-$, due to PH\ symmetry
the final BdG states $\left \vert \psi_{\mathrm{f}}\right \rangle $ turn into $(%
\begin{array}
[c]{c}%
e^{i\varphi}\left \vert +\right \rangle \\
e^{-i\varphi}\left \vert -\right \rangle
\end{array}
).$ Here $\phi_{1},$ $\phi_{2}$ and $\varphi$ are real numbers to be
determined. From the definition of $(%
\begin{array}
[c]{c}%
\left \vert +\right \rangle \\
\left \vert -\right \rangle
\end{array}
)=(%
\begin{array}
[c]{c}%
\left \vert \gamma_{1}\right \rangle -i\left \vert \gamma_{2}\right \rangle \\
\left \vert \gamma_{1}\right \rangle +i\left \vert \gamma_{2}\right \rangle
\end{array}
)/\sqrt{2}$, we can obtain two solutions: one is $\phi_{1}=\pi,$ $\phi_{2}=0,$
$\varphi=-\frac{\pi}{2},$ the other is $\phi_{1}=0,$ $\phi_{2}=\pi,$
$\varphi=\frac{\pi}{2}$. In this paper, due to equivalence property, we focus
on the first solution. As a result, on the basis of $(%
\begin{array}
[c]{c}%
\left \vert +\right \rangle \\
\left \vert -\right \rangle
\end{array}
)$, the braiding operator for BdG states becomes $\mathcal{R}=e^{-i\pi
\sigma^{z}/2}$. If we consider the initial BdG states to be $\left \vert
\psi_{\mathrm{i}}\right \rangle =(%
\begin{array}
[c]{c}%
\left \vert \gamma_{1}\right \rangle \\
\left \vert \gamma_{2}\right \rangle
\end{array}
)$, after the braiding operation, the final BdG states are $(%
\begin{array}
[c]{c}%
-\left \vert \gamma_{2}\right \rangle \\
\left \vert \gamma_{1}\right \rangle
\end{array}
).$

In addition, we can represent the braiding operation for BdG states by an
"unitary" transformation $\mathcal{R}_{ij}=e^{\frac{\pi}{2}\gamma_{i}%
\gamma_{j}}=\gamma_{i}\gamma_{j}$. For three MFs $\gamma_{i}$, $\gamma_{j}$,
$\gamma_{k}$, we found an anti-commutating relation as $\{ \mathcal{R}_{ij},$
$\mathcal{R}_{jk}\}=0.$ As a result, we can check the "statistics" of BdG
states to verify the non-Abelian statistics of MFs. In particular, the direct
relationship between the braiding operation $U_{ij}$ and $\mathcal{R}$ is
\[
\gamma_{k}\rightarrow \gamma_{k}^{\prime}=U_{ij}^{\dagger}\gamma_{k}%
U_{ij}=\mathcal{R}_{ij}\gamma_{k}.
\]

Based on two typical topological SCs, 1D $p$-wave topological SC and 2D
$p_{x}+ip_{y}$-wave topological SC, we verify the non-Abelian statistics of
emergent MFs in SCs by simulating the braiding processes in BdG representation
numerically. To characterize the braiding process, we introduce two
parameters, the amplitude of the function-overlap $\mathcal{O}=\left \vert
\left \langle \psi_{\mathrm{i}}\right \vert \psi_{\mathrm{f}}\rangle \right \vert
$ and the relative Berry phase $\Phi=\left \vert \varphi_{1\mathrm{B}}%
-\varphi_{0\mathrm{B}}\right \vert $. If we get $\mathcal{O}=1$ and $\Phi=\pi$,
MFs obey non-Abelian statistics. Therefore, we can distinguish the statistics
for the MFs by calculating $\mathcal{O}$ and $\Phi$.

The first model is 1D $p$-wave SC on T-junction, which consists of two parts,
a "$-$"-line (A$\leftrightarrow$C, of which the length is $L_{-}$) and a
"$\mid$"-line (D$\leftrightarrow$B of which the length is $L_{\mid}$). See the
illustration in Fig.1.(a). The pair order parameter on "$-$"-line is $\Delta$
and the pair order parameter on "$\mid$"-line is $i\Delta$. The Hamiltonian of
the system can be written as $H=H_{-}+H_{\mid}+H_{D}$ where
\begin{align}
H_{-}  &  =-J\sum \limits_{j=1}^{L_{-}-1}c_{j+1}^{\dagger}c_{j}+\Delta
\sum \limits_{j=1}^{L_{-}-1}c_{j+1}^{\dagger}c_{j}^{\dagger}+h.c.-\sum
\limits_{j=1}^{L_{-}}\mu_{j}c_{j}^{\dagger}c_{j},\nonumber \\
H_{\mid}  &  =-J\sum \limits_{j=1}^{L_{\mid}-1}a_{j+1}^{\dagger}a_{j}%
+i\Delta \sum \limits_{j=1}^{L_{\mid}-1}a_{j+1}^{\dagger}a_{j}^{\dagger
}+h.c-\sum \limits_{j=1}^{L_{\mid}}\mu_{j}a_{j}^{\dagger}a_{j},\nonumber \\
H_{D}  &  =Ja_{j=1}^{\dagger}c_{j_{D}}+i\Delta a_{j=1}^{\dagger}c_{j_{D}%
}^{\dagger}+h.c.
\end{align}
where $c_{j}$ ($a_{j}$) is the annihilation operator of spinless fermions on
"$-$($\mid$)"-line and $j_{D}$ denotes the touching point of "$\mid$"-line on
"$-$"-line (D point in Fig.1.(a)). $J$ is the hopping strength, $\Delta$ is
the SC pairing order parameter and $\mu_{i}$ is the on-site chemical
potential, respectively. In the followings, we choose $\Delta=J$ and the
lattice constant is set to be unit.

\begin{figure}[ptbh]
\includegraphics[width=0.5\textwidth]{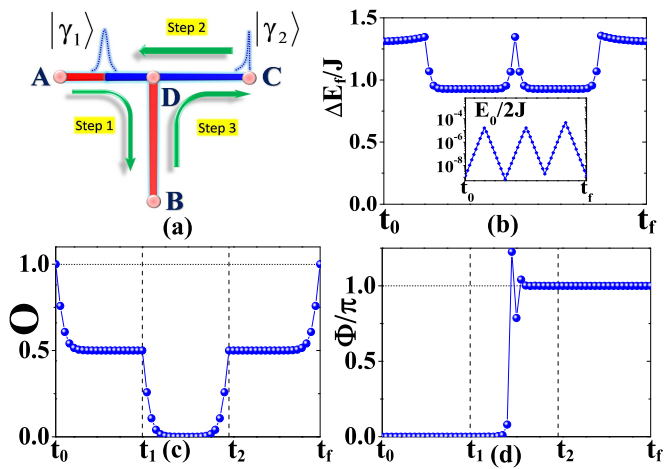}\caption{(color online) (a)
Illustration of T-junction and T-shape braiding path. We denote the
topological $p$-wave SC by blue line and non-topological $p$-wave SC by red
line. As a result, there always exist a pair of MFs ($\gamma_{1}$, $\gamma
_{2}$) at two ends of blue lines; (b) The energy gap of the system, $\Delta
E_{f}$. The inset is the energy $E_{0}$ of Majorana state $\left \vert
0\right \rangle $; (c) The function-overlap $\mathcal{O}=\left \vert
\left \langle \psi_{\mathrm{i}}\right \vert \psi_{\mathrm{f}}\rangle \right \vert
$ of the Majorana modes during the adiabatic evolutions (the dotted line
denotes $1$); (d) The relative Berry phase $\Phi$ (the dotted lines denotes
$\pi$). In all these figures, we choosing the parameters as $\Delta=J.$}%
\end{figure}

For the case of $\left \vert \mu \right \vert <2t$, the ground state is a
topological SC; for the case of $\left \vert \mu \right \vert >2t$, the ground
state is a non-topological SC. In Fig.1.(a), we denote the topological
$p$-wave SC by blue line and non-topological $p$-wave SC by red line. As a
result, there always exist a pair of MFs ($\gamma_{1}$, $\gamma_{2}$) at two
ends of topological $p$-wave SC (a blue line). For the initial state
($\mathrm{t}=\mathrm{t}_{0}$), we set $\mu=-0.7J$ on "$-$"-line and $\mu=-10J$
on "$\mid$"-line. Thus, the $p$-wave SC on "$-$"-line is topological; while
the $p$-wave SC on "$\mid$"-line is non-topological. By numerical calculations
on T-junction with $L_{-}=20$, $L_{\mid}=10$, we find that there are the BdG
states of two MFs $(\left \vert \gamma_{1}\right \rangle ,$ $\left \vert
\gamma_{2}\right \rangle )$ near A and C, respectively.

In the following parts we will show how to braid the two MFs ($\gamma_{1}$,
$\gamma_{2}$) at two ends of "$-$"-line. As shown in Fig.1.(a), we can move
$\gamma_{1}$ away from A by tuning the on-site chemical potential on T-shape
lattice. Thus,\ braiding two MFs ($\gamma_{1}$, $\gamma_{2}$) is a three-step
process: step 1 is to move MF $\gamma_{1}$ from A to B through D as
A$\rightarrow$D$\rightarrow$B during the time period $\mathrm{t}%
_{1}-\mathrm{t}_{0}$; step 2 is to move $\gamma_{2}$ from C to A through D
during the time period $\mathrm{t}_{2}-\mathrm{t}_{1}$; step 3 is to move
$\gamma_{1}$ from B to C through D during the time period $\mathrm{t}%
_{f}-\mathrm{t}_{2}$.

Then, we adiabatically braid the MFs step-by-step by choosing the T-shape
path. During the braiding process, the minimum value of energy gap $\Delta
E_{f}$ is about $0.9J$ that protected the topological properties of the system
and the stability of the MFs and the maximum value of the energy
$E_{0}=\mathcal{J}/2$ ($E_{1}=-\mathcal{J}/2$) of Majorana state $\left \vert
+\right \rangle $ ($\left \vert -\right \rangle $) is about $10^{-5}J$ that may
lead to a small dynamical phase. After calculating $\mathcal{T}\{ \exp[i%
{\displaystyle \int \nolimits_{t_{0}}^{t_{f}}}
H(t)dt]\left \vert \psi_{\mathrm{i}}\right \rangle \}$, we derive the dynamical
phase and Berry phase of the Majorana zero modes during the braiding process.
Here, $\mathcal{T}$ is the time-ordered-product operator. To guarantee the
adiabatic condition, $\Delta \mathrm{t}\gg \hbar/\Delta E_{f}$, the time period
$\Delta \mathrm{t}$ for moving an MF one lattice constant is very large,
$\Delta \mathrm{t}=10000\hbar/J.$ The total time period for the braiding
operation is $\mathrm{t}_{f}-\mathrm{t}_{0}=58\Delta \mathrm{t}$ and
$\mathrm{t}_{1}-\mathrm{t}_{0}=\mathrm{t}_{2}-\mathrm{t}_{1}=19\Delta
\mathrm{t}$, $\mathrm{t}_{f}-\mathrm{t}_{2}=20\Delta \mathrm{t}$.

The final results are given in Fig.1.(c) and Fig.1.(d). For the initial state
$\left \vert \psi_{\mathrm{i}}\right \rangle $ to be $\left \vert +\right \rangle
$ (or $\left \vert -\right \rangle $), after the braiding operation, we get
$\Phi=\pi$ that means the Berry phases $\varphi_{1\mathrm{B}}$ ($\varphi
_{0\mathrm{B}}$) is $-\pi/2$ (or $\pi/2$). From Fig.1.(d), one can see that
the relative Berry phase $\Phi$ changes abruptly during MF $\gamma_{2}$
crossing D point. The fidelity $\left \vert \left \langle \psi_{\mathrm{i}%
}\right \vert \psi_{\mathrm{f}}\rangle \right \vert $ is very close to $100\%$.
As a result, we verify the non-Abelian-statistics of MFs in 1D p-wave
topological SC system .

The second model is $p_{x}+ip_{y}$-wave SC, of which the Hamiltonian can be
written as\cite{Read00}%
\begin{align}
H  &  =-\sum \limits_{j}\sum \limits_{\widehat{\mu}=\widehat{x},\widehat{y}%
}J_{j}(c_{j+\widehat{\mu}}^{\dagger}c_{j}+c_{j-\widehat{\mu}}^{\dagger}%
c_{j})/2-\mu \sum \limits_{j}c_{j}^{\dagger}c_{j}\nonumber \\
&  +\sum \limits_{j}[\Delta_{j}(c_{j+\widehat{x}}^{\dagger}c_{j}^{\dagger
}+ic_{j+\widehat{y}}^{\dagger}c_{j}^{\dagger})/2+h.c.],
\end{align}
where $c_{j}^{\dagger}/c_{j}$ is the creation/annihilation operator of
spinless fermion. $J_{j}$ is the hopping strength, $\Delta_{j}$ is the SC
pairing order parameter and $\mu$ is the chemical potential, respectively. In
the followings, we choosing the parameters as $\left \vert \Delta
_{j}\right \vert =\left \vert J_{j}\right \vert =J,$ $\mu=J$. Now the ground
state is a topological SC with non-zero Chern number. The lattice constant is
also set to be unit.

In the topological phase of $p_{x}+ip_{y}$-wave SC, we study two MFs
($\gamma_{1}$, $\gamma_{2}$) around two quantized vortices ($\pi$-fluxes) by
numerical calculations on a $20\times20$ lattice. The dotted line denotes the
phase branch-cut of the two $\pi$-fluxes (we call it A-B-C $\pi$-phase
string), along which the hopping parameters and the pairing order
parameters\ change sign, $J_{j}\rightarrow-J_{j},$ $\Delta_{j}\rightarrow
-\Delta_{j}$. Thus, there exists a Majorana zero mode around each end of the
$\pi$-phase string. See the particle density distribution of Majorana modes
$(\left \vert \gamma_{1}\right \rangle ,$ $\left \vert \gamma_{2}\right \rangle )$
around $\pi$-fluxes in Fig.2.(a).

To braid the two MFs ($\gamma_{1}$, $\gamma_{2}$), we choose a $\square$-shape
path rather than traditional T-shape path. As shown in Fig.2.(b),\ we
anticlockwise move two MFs ($\gamma_{1}$, $\gamma_{2}$) through a two-step
process: step 1 is to move MF $\gamma_{1}$ from A to B together with moving MF
$\gamma_{2}$ from C to D during the time period $\mathrm{t}_{1}-\mathrm{t}%
_{0}$; step 2 is to move MF $\gamma_{1}$ from B to C together with moving MF
$\gamma_{2}$ from D to A during the time period $\mathrm{t}_{f}-\mathrm{t}%
_{1}$. However, after the two-step braiding process, the A-B-C $\pi$-phase
string changes into an C-D-A $\pi$-phase string. Thus, to return the initial
configuration, we have to deform the C-D-A $\pi$-phase string to A-B-C $\pi
$-phase string by doing local Z2 transformation on the fermion fields inside
the $\square$-shape closed loop A-B-C-D, $c_{j}\rightarrow-c_{j}$. Eventually,
we adiabatically do the braiding operation.

\begin{figure}[ptbh]
\includegraphics[width=0.5\textwidth]{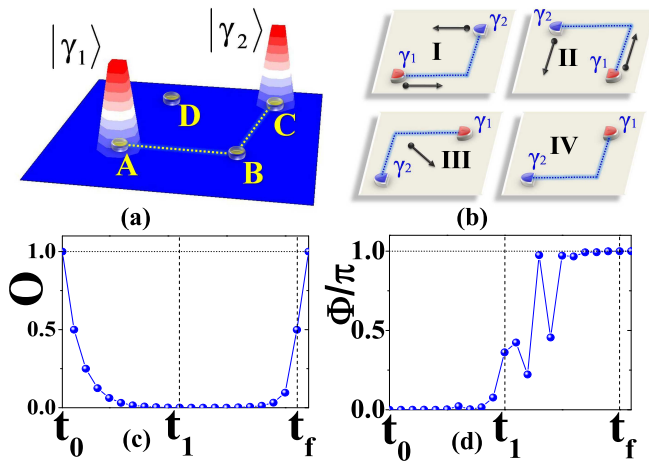}\caption{(color online) (a) The
particle density distribution of Majorana modes ($\left \vert \gamma
_{1}\right \rangle ,$ $\left \vert \gamma_{2}\right \rangle $) around two $\pi
$-fluxes in $p_{x}+ip_{y}$ topological superconductor. The dotted line denotes
A-B-C $\pi$-phase string that connects the two MFs; (b) The $\square$-shape
path of two-step braiding process: I$\rightarrow$II ($\mathrm{t}%
_{0}<\mathrm{t}<\mathrm{t}_{1}$) and II$\rightarrow$III ($\mathrm{t}%
_{1}<\mathrm{t}<\mathrm{t}_{f}$). III$\rightarrow$IV denotes the string
deformation; (c) The function-overlap $\mathcal{O}=\left \vert \left \langle
\psi_{\mathrm{i}}\right \vert \psi_{\mathrm{f}}\rangle \right \vert $ of the
Majorana modes during the adiabatic evolutions (the dotted line denotes $1$);
(d) The relative Berry phase $\Phi$ (the dotted lines denotes $\pi$). In all
these figures, we choosing the parameters as $\Delta=J,$ $\mu=J$. The last
spots of the data in (c) and (d) are obtained from string deformation rather
than braiding operations. }%
\end{figure}

Then we derive the Berry phases of the Majorana states $\left \vert
+\right \rangle $\ and $\left \vert -\right \rangle $ by calculating
$\mathcal{T}\{ \exp[i%
{\displaystyle \int \nolimits_{t_{0}}^{t_{f}}}
H(t)dt](%
\begin{array}
[c]{c}%
\left \vert +\right \rangle \\
\left \vert -\right \rangle
\end{array}
)\}.$ In numerical calculations, the time period for moving an MF one lattice
constant is $\Delta \mathrm{t}=400\hbar/J$ (or the total time period for the
braiding operation is $\mathrm{t}_{f}-\mathrm{t}_{0}=20\Delta \mathrm{t}$ and
$\mathrm{t}_{1}-\mathrm{t}_{0}=\mathrm{t}_{f}-\mathrm{t}_{1}=10\Delta
\mathrm{t}$). During the braiding operation, the energy gap of the system is
always $\Delta E_{f}\equiv2J$ and the energy splitting between two BdG states
of Majorana modes is extremely tiny, $\max \mathcal{J}<10^{-7}J$. As a result,
the dynamical phases of the Majorana states $\left \vert 1\right \rangle $\ and
$\left \vert 0\right \rangle $ are about $\pm10^{-5}\pi$ that are too small to
cause errors to Berry phase calculations.

The results of the amplitude of the function-overlap $\mathcal{O}=\left \vert
\left \langle \psi_{\mathrm{i}}\right \vert \psi_{\mathrm{f}}\rangle \right \vert
$ and those of the relative Berry phase $\Phi=\left \vert \varphi_{1\mathrm{B}%
}-\varphi_{0\mathrm{B}}\right \vert $ are given in Fig.2.(c) and Fig.2.(d),
respectively. From Fig.2.(d), one can see that the relative Berry phase is
about $\Phi=0.9998\pi$ which really closes to $\pi$. The fidelity
$\mathcal{O}=\left \vert \left \langle \psi_{\mathrm{i}}\right \vert
\psi_{\mathrm{f}}\rangle \right \vert $ is up to $99.98\%$. As a result, MFs
induced by the vortices in $p_{x}+ip_{y}$ topological SC obey non-Abelian statistics.

Finally, we draw the conclusion. In this paper we found that the non--Abelian
statistics of MFs can be represented by the anti-commutating relation of BdG
states ($\{ \mathcal{R}_{ij},$ $\mathcal{R}_{jk}\}=0$). From the relationship
between the braiding operator of MFs and that of BdG states, we develop an
numerical method to verifying non-Abelian statistics of MFs in 1D and 2D
topological SCs. The numerical results exactly confirm our prediction.

\begin{center}
{\textbf{* * *}}
\end{center}

This work is supported by National Basic Research Program of China (973
Program) under the grant No. 2011CB921803, 2012CB921704 and NSFC Grant
No.11174035, 11474025, 11404090 and SRFDP, the Fundamental Research Funds for
the Central Universities.


\begin{thebibliography}{99}                                                                                               %


\bibitem {Majorana}E. Majorana, Soryushiron Kenkyu \textbf{63} 149 (1981).

\bibitem {Wilczek}F. Wilczek, Nature Phys. \textbf{5} 614 (2009).

\bibitem {Martin}M. Leijnse and K. Flensberg, arXiv:1206.1736.

\bibitem {Kitaev01}A.Y.\ Kitaev, Phys.\ Usp.\  \textbf{44}, 131 (2001).

\bibitem {Fu08}L.\ Fu and C.L.\ Kane, Phys.\ Rev.\ Lett.\  \textbf{100}, 096407 (2008).

\bibitem {Lutchyn10}R. M.\ Lutchyn, et.al, Phys.\ Rev.\ Lett.\  \textbf{105},
077001 (2010).

\bibitem {Oreg10}Y.\ Oreg, et.al, Phys.\ Rev.\ Lett.\  \textbf{105}, 177002 (2010).

\bibitem {Sau10}J.\ D.\ Sau, et.al, Phys.\ Rev.\ Lett.\  \textbf{104}, 040502 (2010).

\bibitem {Alicea10}J.\ Alicea, Phys.\ Rev.\ B \textbf{81}, 125318 (2010).

\bibitem {Potter10}A.\ C.\ Potter and P.\ A.\ Lee,
Phys.\ Rev.\ Lett.\  \textbf{105}, 227003 (2010).

\bibitem {Alicea11}J.\ Alicea, et.al, Nature Phys.\  \textbf{7}, 412 (2011).

\bibitem {Halperin12}B.\ I.\ Halperin, et.al, Phys.\ Rev.\ B \textbf{85},
144501 (2012).

\bibitem {Stanescu13}T.\ D.\ Stanescu and S.\ Tewari, J.\ Phys. C \textbf{25},
233201 (2013).

\bibitem {Mourik12}V.\ Mourik, et.al, Science \textbf{336}, 1003 (2012).

\bibitem {Das12}A.\ Das, et.al, Nature Phys. \textbf{8}, 887 (2012).

\bibitem {Deng12}M.\ T.\ Deng, et.al, Nano Lett.\  \textbf{12}, 6414 (2012).

\bibitem {Rokhinson12}L.\ P.\ Rokhinson, et.al, Nat.\ Phys.\  \textbf{8}, 795 (2012).

\bibitem {Churchill13}H.\ O.\ H.\ Churchill, et.al, Phys.\ Rev.\ B
\textbf{87}, 241401(R) (2013).

\bibitem {ki2}A. Kitaev, Ann. Phys. \textbf{321}, 2 (2006).

\bibitem {sar}C. Nayak, et.al, Rev. Mod. Phys. \textbf{80}, 1083 (2008).

\bibitem {free}M. H. Freedman, et.al, Math. Phys. \textbf{227}, 605 (2002).

\bibitem {sar1}S. Das Sarma, et.al, Phys. Rev. Lett. \textbf{94}, 166802 (2005).

\bibitem {ge}L. S. Georgiev, Phys. Rev. \textbf{B 74}, 235112 (2006); L. S.
Georgiev, Nucl. Phys. B\textbf{ 789}, 552 (2008).

\bibitem {Read00}N.\ Read and D.\ Green, Phys.\ Rev.\ B \textbf{61}, 10267 (2000).

\bibitem {Ivanov01}D.\ A.\ Ivanov, Phys.\ Rev.\ Lett.\  \textbf{86}, 268 (2001).

\bibitem {Nayak08}C.\ Nayak, et.al, Rev.\ Mod.\ Phys.\  \textbf{80}, 1083 (2008).

\bibitem {Alicea12}J.\ Alicea, Rep.\ Prog.\ Phys.\  \textbf{75}, 076501 (2012).

\bibitem {kou}Y. J. Wu, et.al, Phys. Rev. A. \textbf{90}, 022324 (2014).

\bibitem {moo}G.\ Moore and N.\ Read, Nucl.\ Phys.\ B \textbf{360}, 362 (1991).

\bibitem {wen}Xiao-Gang Wen, Phys. Rev. Lett. \textbf{66}, 802 (1991).

\bibitem {nay}C. Nayak, \& F. Wilczek, Nucl. Phys. B\textbf{ 479}, 529-553 (1996).

\bibitem {free1}M. Freedman, et.al, Phys. Rev. B \textbf{73}, 245307 (2006).
\end{thebibliography}
\end{document}